\documentclass{ws-ijmi}

\usepackage{amsthm}
\usepackage{algpseudocode}
\usepackage[plain]{algorithm}
\usepackage{amssymb, amsmath, array}
\usepackage{balance, booktabs, bm}
\usepackage[labelsep=period]{caption}
\usepackage{calc, color, comment}
\usepackage{cases}
\usepackage{empheq, etoolbox}
\usepackage{flafter, float}
\usepackage{graphicx}
\usepackage{epsfig, epstopdf}
\usepackage{latexsym, lipsum, lineno}
\usepackage{mathtools, makecell, multirow, multicol}
\usepackage{pdflscape, pict2e}
\usepackage{rotating}
\usepackage{setspace, subcaption}
\usepackage{tabularx}
\usepackage[flushleft]{threeparttable}
\usepackage{wrapfig}
\usepackage[dvipsnames]{xcolor}
\usepackage{xtab}
\usepackage{url, overcite}

\newtheorem*{definition*}{Definition}

\newtheorem*{remark*}{Remark}
\newtheorem{remark}{Remark}

\begin{document}

%%% TYPE OF ARTICLE 
\articletype{Original Article}

%%% PAPER'S INFORMATION
\catchline{00}{0}{0000}{article id}{}

%%% PRELIMINARY INFORMATION
\markboth
{Aditya Dewanto Hartono, T\^{o}n Vi\d {\^{e}}t T\d{a}, Linh Thi Hoai Nguyen}
{Geometrical Structures for Predator-Avoidance Fish School}

%%% TITLE OF THE PAPER
\title{A Geometrical Structure for Predator-Avoidance Fish Schooling}

%%% AUTHORS
\author{
Aditya Dewanto Hartono$^\dag$,
T\^{o}n Vi\d {\^{e}}t T\d{a}$^\dag$, 
and Linh Thi Hoai Nguyen$^\ddag$}

%%% AUTHORS' AFFILIATIONS
\address{
\small
$^\dag$Mathematical Modeling Laboratory, Department of Agro-environmental Sciences, Kyushu University, Japan\\[1pt]
$^\ddag$Institute of Mathematics for Industry, Kyushu University, Japan\\[1pt]
}

\maketitle

%%% ABSTRACT OF THE PAPER
\begin{abstract}
This paper conducts a numerical study of a geometrical structure called $\epsilon$-school for predator-avoidance fish schools, based on our previous mathematical model. Our results show that during a predator attack, the number of $\epsilon$-school increases from one to a certain value. After the attack, the number of $\epsilon$-school decreases in the first two predator-avoidance patterns, but continues to increase in the third pattern.  A constant value for the number of the $\epsilon$-school is observed in the last pattern.  These suggests that when the predator is approaching, each individual in the school focuses more on avoiding the predator, rather than on interacting with its schoolmates. Such a trait is in agreement with real-life behavior in the natural ecosystem.
\end{abstract}


\vspace*{-0.5cm}

\begin{multicols}{2}

%%% SECTION 1: INTRODUCTION
\section{Introduction}
Fish schooling is a remarkable phenomenon in the aquatic world that has captivated many researchers. The synchronized movement of hundreds or even thousands of fish in a school is a complex and highly organized trait. Such unique swarm behavior has been the subject of numerous studies in various disciplines, including biology, physics, and mathematics\cite{Aoki1982School,Camazine2001Biological,Cucker2007Mathematics,Reynolds1987Flocks,Couzin2009Collective,Vicsek2012CollectiveMotion}.

Studying fish schooling from a mathematical point of view is important. We can gain insights into the rules governing the behavior of individuals  and have a deeper understanding of the underlying patterns and dynamics of collective behavior in animal groups. Mathematical models can make predictions and analyze the effects of various factors on the behavior of a school of fish, such as the interaction between individual fish, environmental conditions, and external stimuli. This information can have important implications for fields such as fisheries management, wildlife conservation, and aquatic ecology. It can also have practical applications, such as in the design of swarm robotics to accomplish tasks that would be difficult or impossible for a single robot to accomplish on its own, and the design of software for autonomous vehicles (e.g. self-driving cars) that use collision-avoidance rule of fish.

We have studied fish schooling from the mathematical point of view for more than a decade. In Ref.~\citen{Uchitane2012ODE}, we constructed a stochastic differential equation model for fish schooling, which is based on the biological interaction rules outlined by Camazine \textit{et al.}\cite{Camazine2001Biological}. A geometrical analysis of such a model is then presented in Ref.~\citen{Linh2015Quantitative}. In Ref.~\citen{Linh2016ObstacleAvoiding}, we investigated the obstacle-avoiding patterns of fish schools by incorporating an obstacle-avoidance rule into our original model of Ref.~\citen{Uchitane2012ODE}. Therein, for the first time, we were able to quantify the cohesiveness of fish schools.

In Ref.~\citen{TonandLinh2018Foraging}, we developed a mathematical model for the foraging behavior of fish schools. Our results revealed that when fish form a unitary formation in terms of school, they are able to locate the food more effectively: such a trait is one of the benefits of constituting a school that is consistent with real-life situation in the natural ecosystem\cite{Pitcher1982FishFood, Pitcher1986Functions, Krause2010SwarmIntelligence, Ioannou2017Swarm}.

In Ref.~\citen{Hartono2022Stochastic}, we proposed a model of stochastic differential equations to describe predator-avoidance behavior of a prey fish school. The model is as follows\cite{Hartono2022Stochastic}:
\begin{equation} \label{eq01}
\begin{aligned}
\begin{cases}
d{x}_{i}(t) = &v_{i}dt + \sigma_{i}d{w}_{i}(t), \quad (i = 1, 2, \dots, N),\\
d{v}_{i}(t) = &\Bigg[
-\alpha \sum \limits_{j = 1, j \ne i}^{N} \left(\frac{r^{p}}{\|x_{i} - x_{j}\|^{p}} - \frac{r^{q}}{\|x_{i} - x_{j}\|^{q}}\right) \\
&\hspace*{0.3cm} \times \left(x_{i} - x_{j}\right)\\
&\hspace*{0.3cm} -\beta \sum \limits_{j = 1, j \ne i}^{N} \left(\frac{r^{p}}{\|x_{i} - x_{j}\|^{p}} + \frac{r^{q}}{\|x_{i} - x_{j}\|^{q}}\right)\\
&\hspace*{0.3cm} \times \left(v_{i} - v_{j}\right) + H(x_{i}, y)
\Bigg]dt, \\
&\hspace*{0.3cm} (i = 1, 2, \dots, N),\\
dy(t) =& vdt + \sigma d{w}_{t},\\
dv(t) =& F\left(x_{i}, v_{i}, y, v\right)dt,
\end{cases}
\end{aligned}
\end{equation}
coupled with a \textquotedblleft being eaten\textquotedblright\ condition.

Here, $N$ is the size of (prey) fish school;  $x_{i}(t)$ and $v_{i}(t)$ $(i = 1, 2, \dots, N)$ respectively denote the position and velocity in $\mathbb R^d \, (d=2, 3)$ of the $i$-th prey fish at time $t$; $y(t)$ and $v(t)$ correspondingly represent the position and velocity of the predator at time $t$; and $\|\cdot\|$ designates the Euclidean norm of a vector.  The \textquotedblleft being eaten\textquotedblright\ condition occurs when the predator is within a distance of $r$ from the $i$-th fish, i.e., $\|y -x_i\|<r$, resulting in the model changing from $N:1$ ($N$ preys, 1 predator) to $(N-1):1$.

The first term in Eq.~(\ref{eq01}) is a stochastic differential equation for the unknown $x_{i}(t)$, where $\sigma_idw_i$ $(i=1, 2 \dots N)$ denotes a stochastic differentiation of $d$-dimensional independent Brownian motion defined in a filtered probability space\cite{Uchitane2012ODE}. The second expression is a deterministic equation for the unknown $v_{i}(t)$. Parameters $1 < p < q < \infty$ are fixed exponents; $\alpha$ and $\beta$ designate positive coefficients of attraction and velocity matching among the individual prey, respectively; and $r > 0$ depicts the critical distance between two individuals in the school.
 
The third expression of Eq.~(\ref{eq01}) is again a stochastic equation for the unknown $y(t)$ in which $w(t)$ is a $d$-dimensional Brownian motion in the same filtered probability space which is independent of $w_i(t), i=1,2,\dots, N$. The last term of Eq.~(\ref{eq01}) is a deterministic equation for the unknown $v(t)$.
 
The function $H(x_{i}, y)$ represents the mechanism adopted by an individual prey fish to avoid the predator. It takes the following remark:
\begin{equation} \label{eq02}
H(x_{i}, y) = \delta \frac{R_{1}^{\theta_{1}}}{\|x_{i} - y\|^{\theta_{1}}} \left(x_{i} - y\right),
\end{equation}
where $R_{1} > r$, $\delta$, and $\theta_{1}$ are positive constants. 

On the other hand, the function $F\left(x_{i}, v_{i}, y, v\right)$ manifests the hunting strategy of the predator. Here, we devised two hunting tactics of the predator, namely (i) the predator attacks the center of the schooling prey (hunting tactic I), and (ii) the predator focuses its attack on the nearest prey (hunting tactic II). The mathematical expressions for each of the prescribed predator's hunting strategies are respectively defined as follows:
\begin{equation} \label{eq03}
\begin{aligned}
F\left(x_{i}, v_{i}, y, v \right) = 
& -\dfrac{R_{2}^{\theta_{2}}}{\|y - x_{c}\|^{\theta_{2}}} \times\\ 
&\Big[
\gamma_{1} \left(y - x_{c}\right) + \gamma_{1} \gamma_{2} \left(v - v_{c}\right)
\Big],
\end{aligned}
\end{equation}
%%%
\begin{equation} \label{eq04}
\begin{aligned}
F\left(x_{i}, v_{i}, y, v \right) = 
&-\dfrac{1}{N} \sum \limits_{j = 1}^{N} \frac{R_{2}^{\theta_{2}}}{\|y - x_{j}\|^{\theta_{2}}} \times\\
&\Big[
\gamma_{1} \left(y - x_{j}\right) + \gamma_{1} \gamma_{2} \left(v - v_{j}\right)
\Big].
\end{aligned}
\end{equation}

In Eq.~(\ref{eq03}), $x_{c}$ and $v_{c}$ respectively denote the center position and velocity of the schooling prey; we defined them as the average value of the positions and velocities of all the individual prey constituting the school:
\begin{equation} \label{eq05}
x_{c} = \dfrac{1}{N} \sum \limits_{i = 1}^{N} x_{i}, \quad
v_{c} = \dfrac{1}{N} \sum \limits_{i = 1}^{N} v_{i}.
\end{equation}
Parameters $R_{2} > r$, $\theta_{2}$, $\gamma_{1}$, and $\gamma_{2}$ are positive constants. Meanwhile, in Eq.~(\ref{eq04}), $x_{j}$ designates the position of each individual prey fish.

In our previous work\cite{Hartono2022Stochastic}, we employed the mathematical model \eqref{eq01} to describe the behavior of the schooling prey fish under attack of a single predator. Therein, we discovered four anti-predation maneuvers of the prey fish school (hereinafter, we also label them as the predator-avoiding patterns). Furthermore, we successfully demonstrated the benefit of constituting a large school size in better escaping the predator's attack. 

There is, however, another crucial aspect of Eq.~(\ref{eq01}) that has not been elaborated further in our earlier work, namely the capability of the model in unveiling the transformation of the geometrical structure of the associated prey school's formation during predation threat of a solitary predator. In this paper, we therefore aim to provide a comprehensive assessment regarding such a central feature of our model. To do so, we introduce the so-called $\epsilon$-school as a mathematical representation of the geometrical structure of the schooling prey's formation. Based on this framework, we undertake numerical simulations to elucidate the transformation of $\epsilon$-school in all the observed four predator-avoidance patterns.
 
The organization of this paper is as follows. In the following section, we provide detailed explanation regarding the notion of $\epsilon$-school and outline the initial conditions for our simulation based on the model \eqref{eq01}. In Section 3, we present the results of the numerical simulations. Lastly, in Section 4, we pose some concluding remarks of the current study.

%%% SECTION 2: EPSILON SCHOOLING 
\section{Preliminary}
In this section, we introduce the concept of $\epsilon$-school and establish initial conditions for our simulations based on the model \eqref{eq01}. The notion of $\epsilon$-school is akin to that of a connected component in an $\epsilon$-graph, as seen in graph theory.

At each time step $t$, we define an $\epsilon$-graph $G(V(t), E(t))$ where the set of vertices 
$$V(t)=\{x_1(t),x_2(t),\ldots,x_N(t)\}$$
 represents the positions of individuals, and the set of edges 
 \begin{align*}
 E(t)=\{ & (x_i(t), x_j(t)) \text{ if } \|x_i(t) - x_j(t)\| \leq \epsilon, \\
 &i, j = 1,2,\ldots,N \}
 \end{align*}
  connects any two individuals whose distance does not exceed $\epsilon$. 

We refer to each connected component of $G(V(t), E(t))$ as an $\epsilon$-school. Furthermore, we denote by $ N_\epsilon(t)$ the number of $\epsilon$-schools in the graph $G(V(t), E(t))$. 

\begin{remark}\label{remark01}
In Ref.~\citen{Linh2015Quantitative}, we introduced a new definition of $\epsilon,\theta$-schooling. The definition states that once the $\epsilon,\theta$-schooling structure has been formed, it will be maintained indefinitely, as long as there are no external factors, such as a predator, that disrupt it. However, in this current paper, the structure changes over time as a result of predator attacks. Therefore, the definition of $\epsilon,\theta$-schooling is not applicable here.
\end{remark}

In this study, we investigate the transformation of the number of $\epsilon$-school structure of the schooling prey fish due to the predator's attack in both two-and three-dimensional spaces $(d = 2, 3)$ for the observed four predator-avoidance patterns in our earlier work\cite{Hartono2022Stochastic}. In all of the simulations, we employ the model \eqref{eq01} with the hunting tactic chosen among Eq.~(\ref{eq03}) and Eq.~(\ref{eq04}), correspondingly. In all of the cases, the number of prey fish is fixed at $N = 40$, and the intensity of noise $\sigma_i=\sigma = 0.01\, (i=1,2,\dots, N)$, while other parameters may vary and are specified as necessary. The maximum simulation time is prescribed at $t_{\textrm{max}} = 3,500$, during which the solitary predator attacks the schooling prey only once. At the beginning of the simulation $(t = 0)$, all prey fish are in an $\epsilon$-school formation, while the predator fish is positioned relatively far from the school.

In the following section, we present the results of the numerical simulations for all the observed four predator-avoidance patterns.

%%% SECTION 3: RESULTS
\section{Results}
As outlined in the Introduction, our primary aim in the present study is to unveil the capability of our predator-prey model in elucidating the transformation of geometrical structure of the prey school's formation during the predator's attack. To do so, we integrate the concept of $\epsilon$-school described in the Preliminary section into the generic model \eqref{eq01}. 

From our previous work\cite{Hartono2022Stochastic}, we obtained four predator-avoidance patterns. We label them as: (i) Pattern I: Split and Reunion, (ii) Pattern II: Split and Separate into Two Groups, (iii) Pattern III: Scattered, and (iv) Pattern IV: Maintain Formation and Distance. In this section, we present the simulation results of each of the observed patterns for two- and three-dimensional simulations, respectively. Let us begin by discussing the simulation results of the two-dimensional cases.

%%% SECTION 3.1: RESULTS IN TWO-DIMENSIONAL SPACE
\subsection{Two-dimensional space}
For the two-dimensional case, the simulations are carried out with the fixed values of $\epsilon = 0.7$. The values of other parameters of the model to obtain each of the predator-avoiding patterns are outlined in Table~\ref{table01}, correspondingly. The associated hunting tactic of the predator for each corresponding patterns is listed in the second column of the table, as well. 

\begin{figure*}[b]
\begin{center}
\includegraphics[width=15cm]{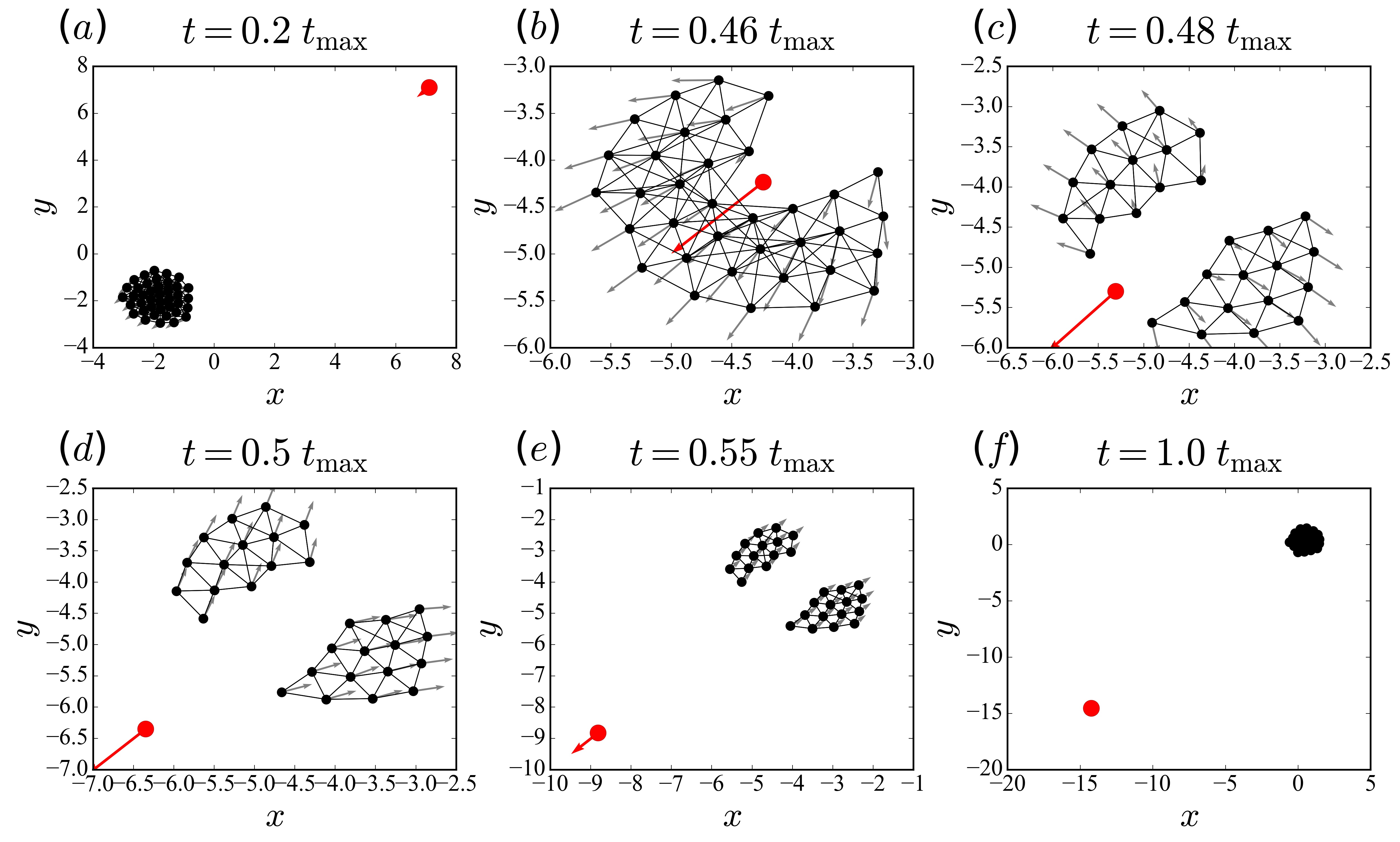}
\end{center}
\caption{
The results of 2D simulation for Pattern I: Split and Reunion. The images demonstrate the behavior of the schooling prey and the predator, as well as the associated condition of $\epsilon$-school at: ($a$) $t = 0.2 t_{\textrm{max}}$, ($b$) $t = 0.46 t_{\textrm{max}}$, ($c$) $t = 0.48 t_{\textrm{max}}$, ($d$) $t = 0.5 t_{\textrm{max}}$, ($e$) $t = 0.55 t_{\textrm{max}}$, and ($f$) $t = t_{\textrm{max}}$, respectively. 
}
\label{fig01}
\end{figure*}

Fig.~\ref{fig01} illustrates the results of the simulation for Pattern I (Split and Reunion), displaying the condition of $\epsilon$-school as simulation time progresses from the early stage until the end of the simulation ($t_{\textrm{max}}$). Therein, the small black dots manifest the schooling prey, while the large red dot designates the predator. The arrow linked to each of the units denotes the direction of movement of that particular unit at the corresponding time. The individuals constituting an $\epsilon$-school (at the corresponding time) are connected to each other through a solid line. Similar configurations apply to the simulation results of other cases.

\begin{table}[H] 
\vspace*{0cm}
\caption{Parameter settings for two-dimensional simulations of predator-avoidance fish schooling.} \label{table01} 
\vspace*{0cm}
\footnotesize \centering
\begin{tabularx}{0.48\textwidth}{
c c c c m{0.01\textwidth}<{\centering} 
m{0.01\textwidth}<{\centering} m{0.02\textwidth}<{\centering} m{0.02\textwidth}<{\centering} 
m{0.02\textwidth}<{\centering} m{0.02\textwidth}<{\centering} 
}
\hline \noalign{\smallskip}
\thead{Pattern\\ID} & Tactic & $\alpha$ & $\beta$ & $\delta$ & $p$ & $\theta_{1}$ & $\theta_{2}$ & $\gamma_{1}$ & $\gamma_{2}$\\
\noalign{\smallskip} \hline \noalign{\medskip}
I & II & 15 & 0.5 & 1 & 4 & 1 & 0.5 & 0.08 & 0.1\\
II & I & 1 & 0.5 & 1 & 4 & 5 & 1 & 0.1 & 0.1\\
III & II & 1 & 0.5 & 5 & 2 & 1 & 2 & 1 & 0.1\\
IV & I & 2 & 0.5 & 0.1 & 2 & 1 & 1 & 5 & 10\\
\noalign{\smallskip} \hline
\end{tabularx}
\end{table}

As can be seen in Fig~\ref{fig01}, as the predator arrives in the vicinity of the prey, the schooling prey reacts accordingly to avoid the predator. Such a maneuver generates a vacuole-form of the schooling prey where each of the individuals tries to get away from the predator (see Fig.~\ref{fig01}($b$)). At this stage, the associated prey still maintains the unitary school formation, as is depicted by the solid lines connecting each of the individuals.

As the predator progresses along its path, a few of the prey fish may be eaten by it. At this stage, all the other \textquotedblleft survived\textquotedblright\ prey responses accordingly by decomposing the unitary formation of the school and temporarily constitutes two smaller groups; each of the groups expands at the right angles away from the direction of the predator's attack. Evidence for this can be seen in Fig.~\ref{fig01}($c$), where two number of $\epsilon$-schools prevail. As the predator moves away from the \textquotedblleft survived\textquotedblright\ prey, the latter entities recombined to form a unitary school formation behind the predator (Fig.~\ref{fig01}($f$)), resulting in the recuperation of the number of $\epsilon$-school into one.

\begin{figure*}[b]
\begin{center}
\includegraphics[width=15cm]{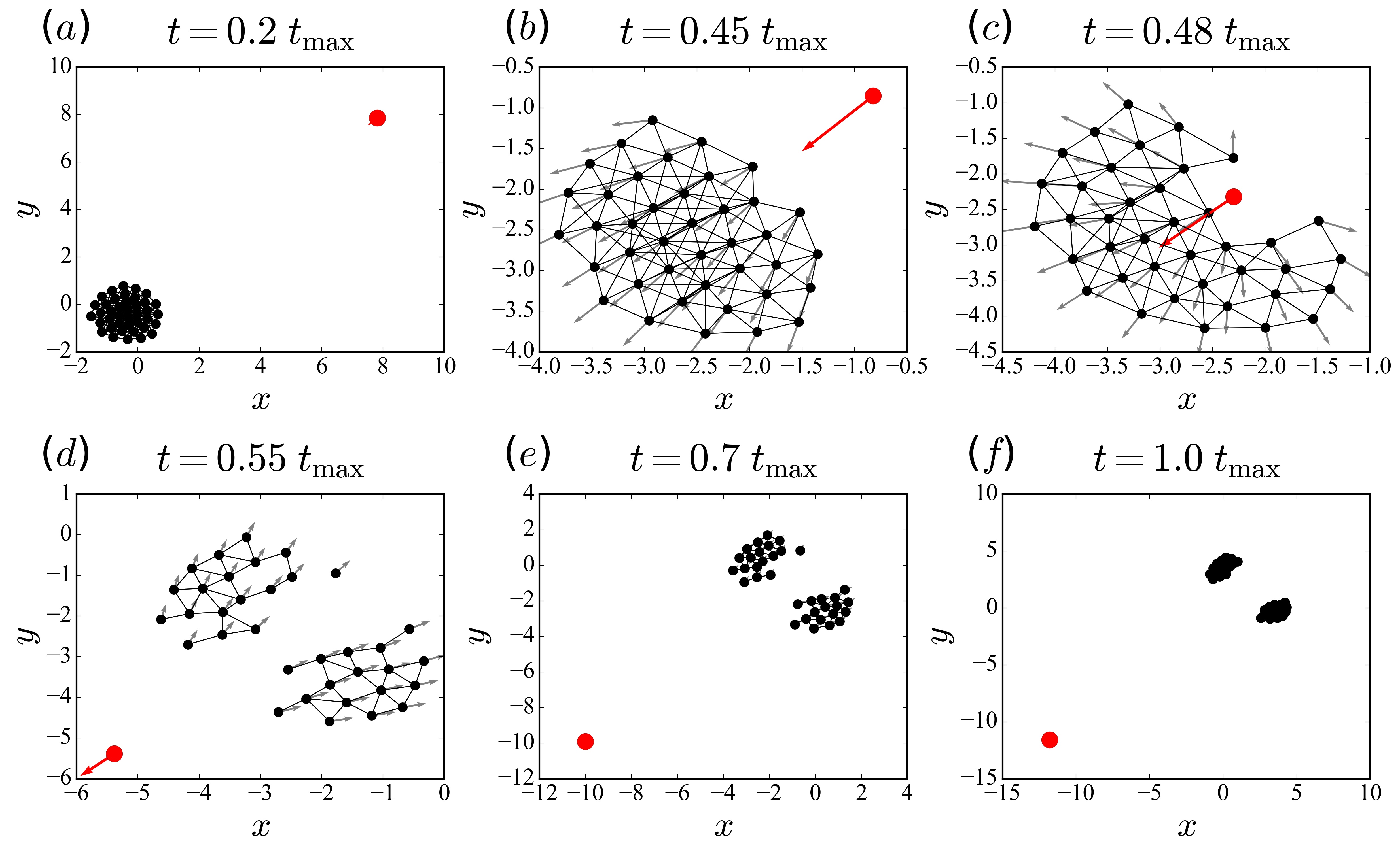}
\end{center}
\caption{
The results of 2D simulation for Pattern II: Split and Separate into Two Groups. The images demonstrate the behavior of the schooling prey and the predator, as well as the associated condition of $\epsilon$-school at: ($a$) $t = 0.2 t_{\textrm{max}}$, ($ b $) $t = 0.45 t_{\textrm{max}}$, ($ c $) $t = 0.48 t_{\textrm{max}}$, ($ d $) $t = 0.55 t_{\textrm{max}}$, ($ e $) $t = 0.7 t_{\textrm{max}}$, and ($ f $) $t = t_{\textrm{max}}$, respectively. 
} 
\label{fig02}
\end{figure*}

Now, let us turn our attention to the next Pattern. Fig.~\ref{fig02} displays the corresponding simulation results for Pattern II (Split and Separate into Two Groups). As can be seen in the figure, the behavior of the schooling prey in Pattern II exhibits similar characteristics with Pattern I during the progression periods of the predator's attack (see Figs.~\ref{fig02}($a$) - ($d$)). The difference, however, is clearly visible in the periods after the attack. Therein, the two smaller groups of the schooling prey do not rejoin into a unitary school formation. Evidence for this is in Figs.~\ref{fig02}($e$) - ($f$). In accordance with this, the number of $\epsilon$-school  decreases to two and remains at that level until the end of the allotted time.

\begin{figure*}[t]
\begin{center}
\includegraphics[width=13cm]{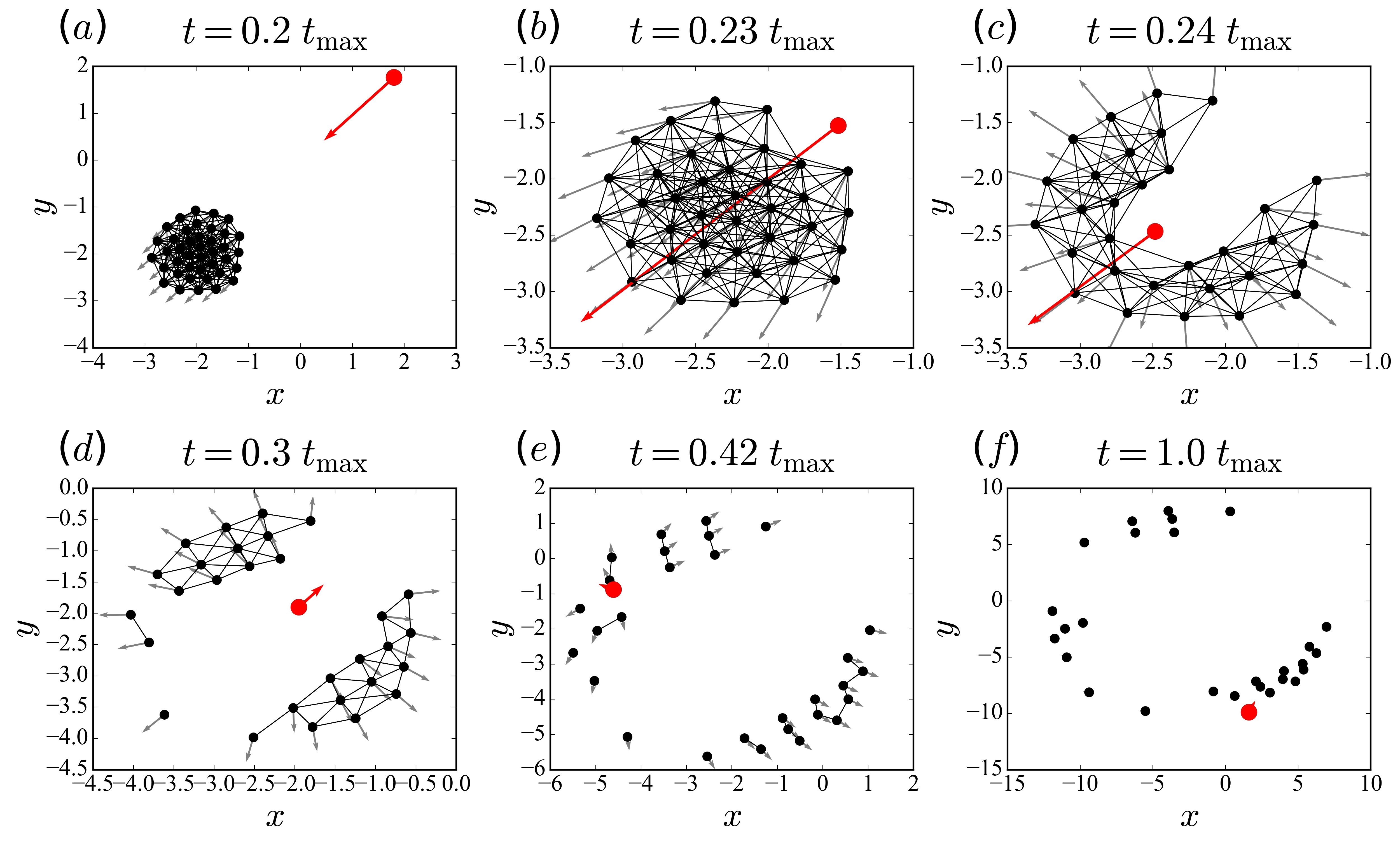}
\end{center}
\caption{
The results of 2D simulation for Pattern III: Scattered. The images demonstrate the behavior of the schooling prey and the predator, as well as the associated condition of $\epsilon$-school at: ($ a $) $t = 0.2 t_{\textrm{max}}$, ($ b $) $t = 0.23 t_{\textrm{max}}$, ($ c $) $t = 0.24 t_{\textrm{max}}$, ($ d $) $t = 0.3 t_{\textrm{max}}$, ($ e $) $t = 0.42 t_{\textrm{max}}$, and ($ f $) $t = t_{\textrm{max}}$, respectively. 
} 
\label{fig03}
\end{figure*}

\begin{figure*}[t]
\begin{center}
\includegraphics[width=13cm]{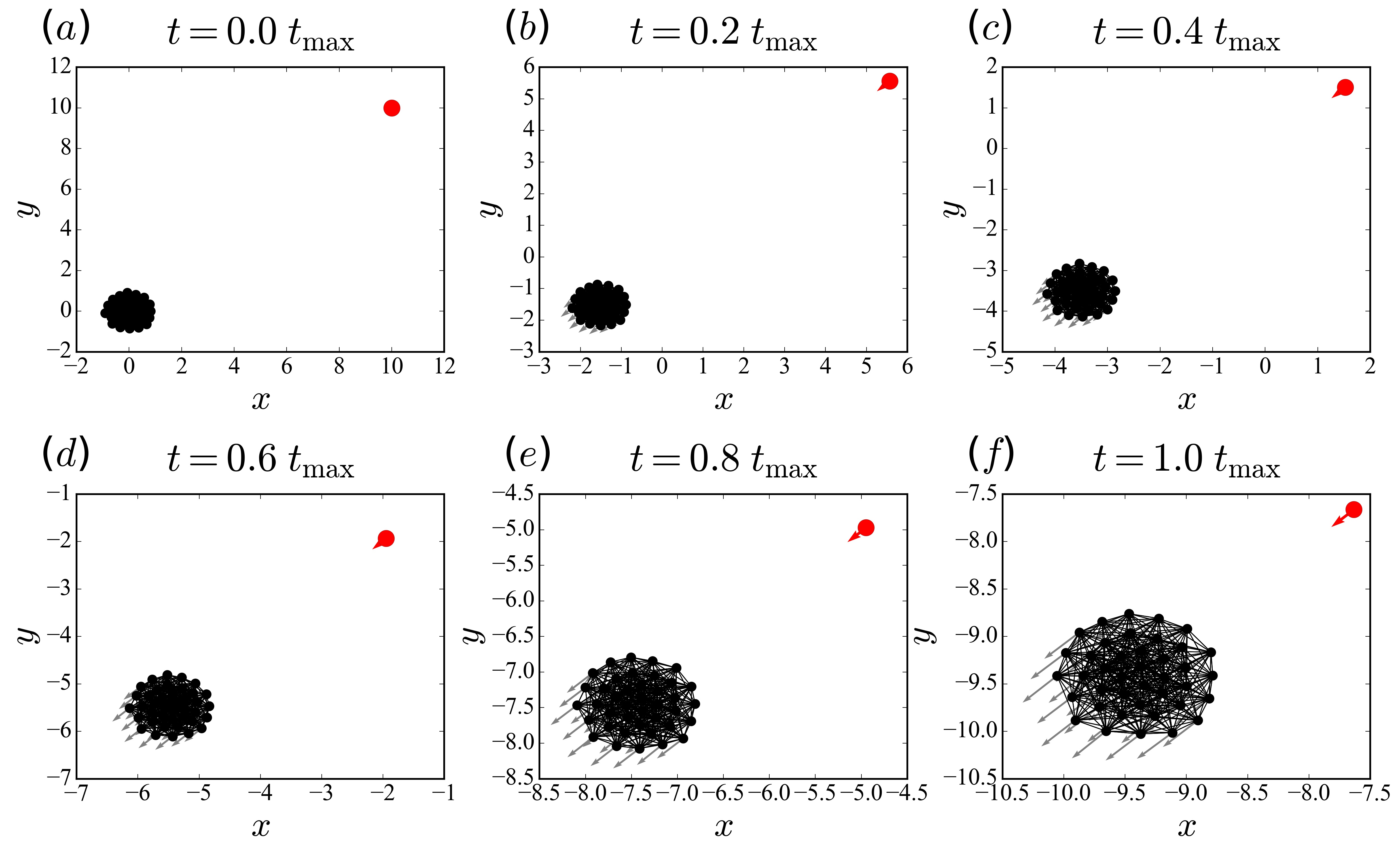}
\end{center}
\caption{
The results of 2D simulation for Pattern IV: Maintain Formation and Distance. The images demonstrate the behavior of the schooling prey and the predator, as well as the associated condition of $\epsilon$-school at: ($ a $) $t = 0$, ($ b $) $t = 0.2 t_{\textrm{max}}$, ($ c $) $t = 0.4 t_{\textrm{max}}$, ($ d $) $t = 0.6 t_{\textrm{max}}$, ($ e $) $t = 0.8 t_{\textrm{max}}$, and ($ f $) $t = t_{\textrm{max}}$, respectively. 
} 
\label{fig04}
\end{figure*}

Next, we move on to Pattern III (Scattered). Here, another distinctive characteristic of the schooling prey appears: the schooling prey seems to display a panic condition and permanently break the unitary school formation as the simulation proceeds. As can be seen in Figs.~\ref{fig03}($b$) - ($e$), as the predator approaches, the number of $\epsilon$-school increases from one into numerous $\epsilon$-schools  based on the number of \textquotedblleft survived\textquotedblright\ prey at the particular time. Many of these structures consist of only one individual prey fish. Because the prey breaks the unitary formation, the predator is in favorable situation to hunt more (available) prey. As a result, the remaining \textquotedblleft survived\textquotedblright\ prey is actively being hunted by the predator, resulting in a continuing panic condition of the individual prey. Such a condition can be identified in Fig.~\ref{fig03}($f$), where numerous structures of $\epsilon$-school prevail as the simulation arrives at $t_{\textrm{max}}$.

For the last anti-predation maneuver (namely Pattern IV: Maintain Formation and Distance), the schooling prey exhibits vigilant behavior to the nearby predator: it maintains a (relatively) safe distance from the predator during the simulation. Consequently, no prey fish is being eaten by the predator. As shown in Fig.~\ref{fig04}, the school of prey fish maintains its unitary $\epsilon$-school  until the end of simulation time.

\begin{figure}[H]
\begin{center}
\includegraphics[width=8.5cm]{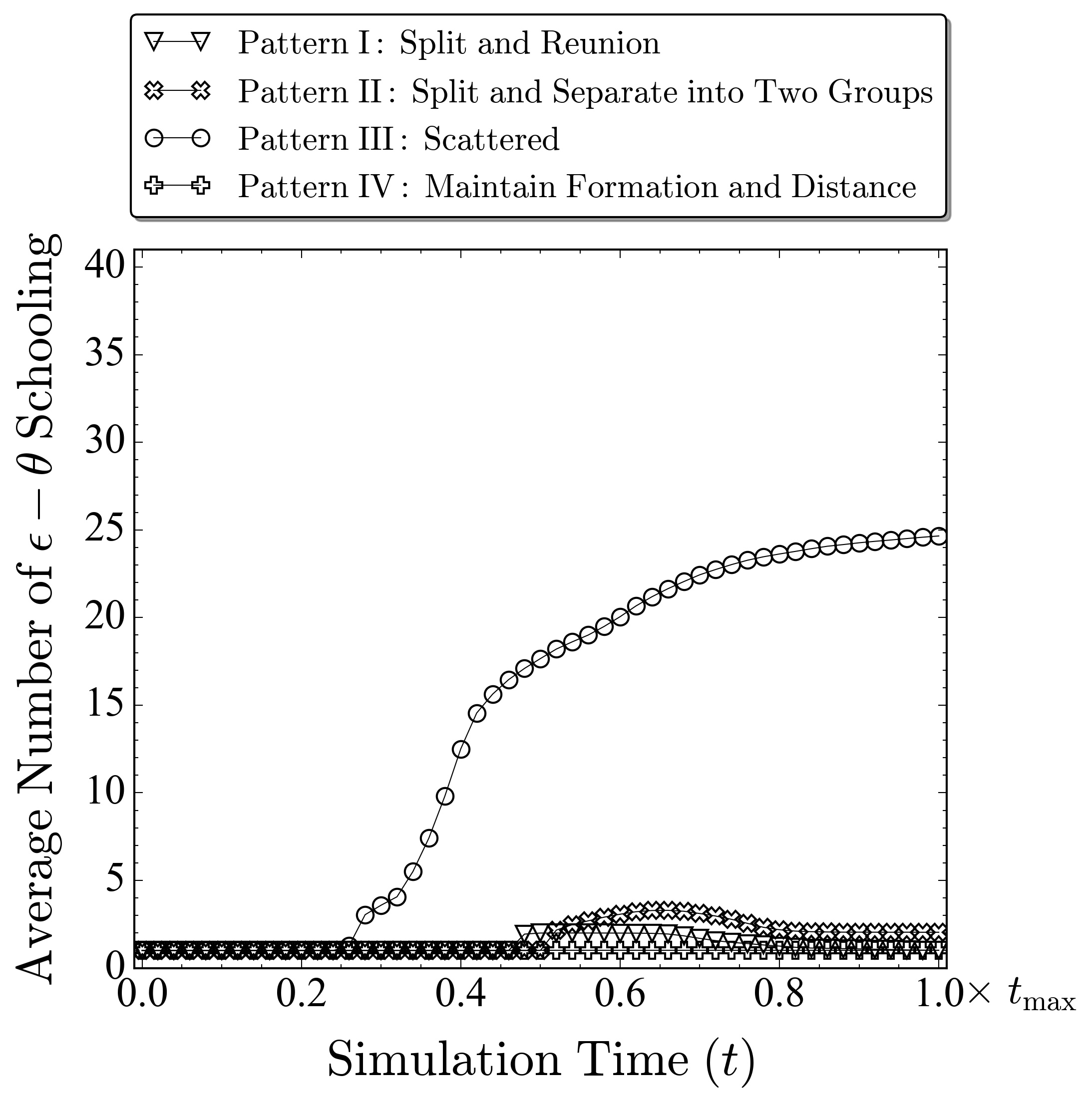}
\end{center}
\caption{Number of $\epsilon$-schools for 2D simulation.}
\label{fig05}
\end{figure}

As our model is stochastic, executing the simulation repeatedly with the same parameters may result in different number of $\epsilon$-schools   each time. Fig.~\ref{fig05} shows the average number of $\epsilon$-schools   at each time step, calculated over 1,000 simulation runs for each pattern, using the same parameters as before. A careful inspection of Fig.~\ref{fig05} reveals that the number of $\epsilon$-schools  increases from one to a certain value, then decrease to one (for Pattern I) or two (for Pattern II). In Pattern III, the number of $\epsilon$-schools   increases as the school becomes more scattered, while in Pattern IV, it remains at one throughout the allotted simulation time. In general, these results suggest that when the schooling prey is under an imminent predation threat, each individual prey fish in the school immediately puts a priority in avoiding predation rather than maintaining their formation with other schoolmates. Such a finding is consistent with real-life observations of the schooling fish\cite{Pavlov2000PatternsSchooling}.

\begin{figure}[H]
\begin{center}
\includegraphics[width=8cm]{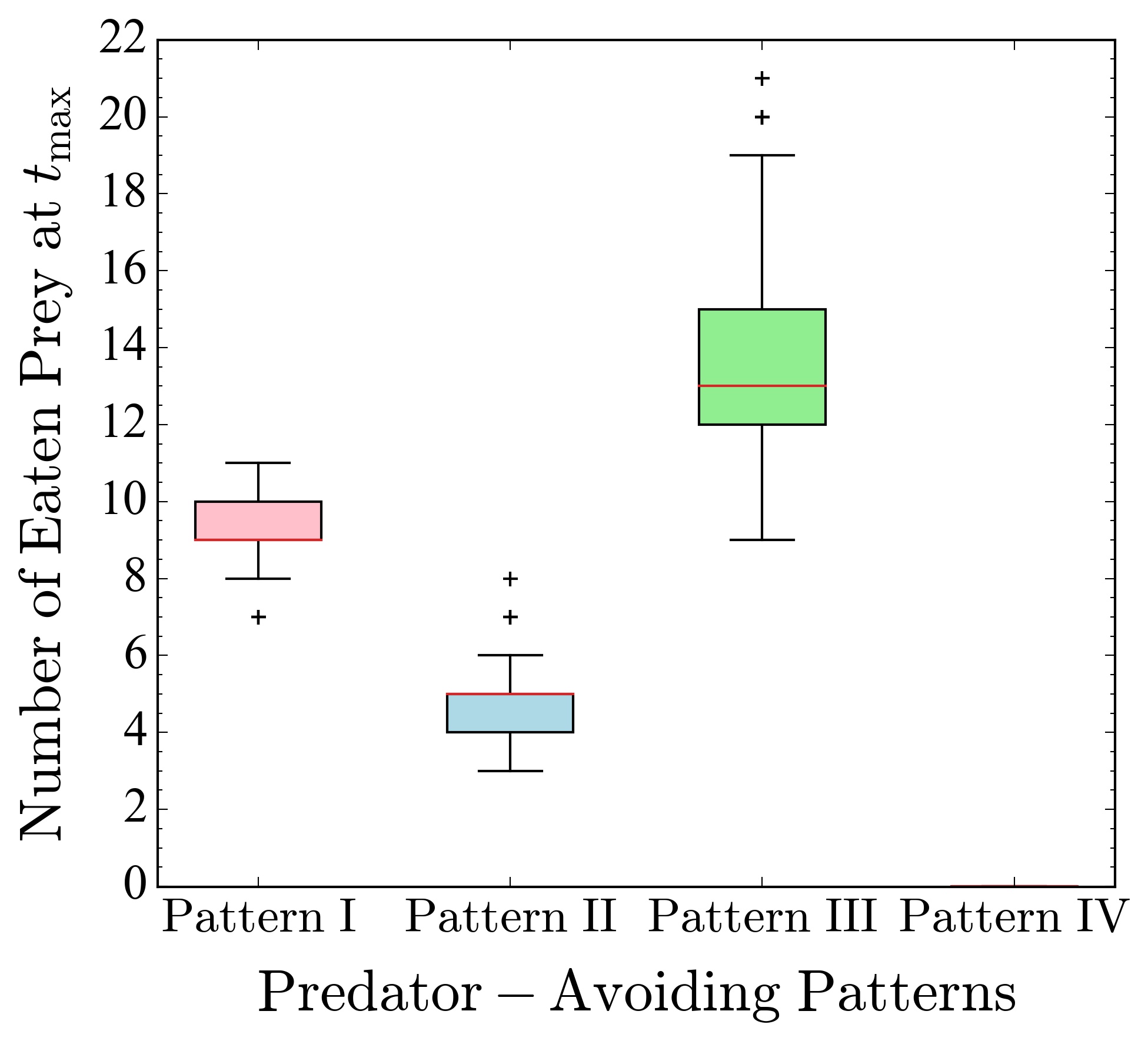}
\end{center}
\caption{Total number of eaten prey for 2D simulation.}
\label{fig06}
\end{figure}

Fig.~\ref{fig06} presents the total number of eaten prey for the two-dimensional simulation space. Here, a total number of 1,000 simulation runs is carried out for each of the corresponding patterns. According to Fig.~\ref{fig06}, it appears that Pattern IV (Maintain Formation and Distance) is the most effective evasive mode for the schooling prey since no single prey is being eaten throughout the simulation runs. On the other hand, the least effective anti-predation mode is displayed by Pattern III (Scattered), with a median of $13$ eaten prey during the predator's attack. Such findings are consistent with observations of diverse fish species in the natural aquatic ecosystem (see, for example, Refs.~\citen{Pavlov2000PatternsSchooling, Partridge1982Structure, Potts1970Schooling, Shaw1978Schooling}).

\vspace*{-0.8cm}

%%% SECTION 3.2: RESULTS IN THREE-DIMENSIONAL SPACE
\subsection{Three-dimensional space}
In this subsection, we present the results of the simulation for the four predator-avoidance patterns alongside their associated $\epsilon$-schools in three-dimensional space. Here, the value of $\epsilon$ is similar to the one employed in the two-dimensional case. Table \ref{table02} summarizes the adopted model parameter settings to run the three-dimensional simulations.

\begin{table}[H] 
\vspace*{0cm}
\caption{Parameter settings for three-dimensional simulations of predator-avoidance fish schooling.} \label{table02} 
\vspace*{0cm}
\footnotesize \centering
\begin{tabularx}{0.48\textwidth}{
c m{0.035\textwidth}<{\centering} c m{0.01\textwidth}<{\centering} m{0.01\textwidth}<{\centering} 
m{0.01\textwidth}<{\centering} m{0.02\textwidth}<{\centering} m{0.02\textwidth}<{\centering} 
m{0.02\textwidth}<{\centering} m{0.02\textwidth}<{\centering} 
}
\hline \noalign{\smallskip}
\thead{Pattern\\ID} & Tactic & $\alpha$ & $\beta$ & $\delta$ & $p$ & $\theta_{1}$ & $\theta_{2}$ & $\gamma_{1}$ & $\gamma_{2}$\\
\noalign{\smallskip} \hline \noalign{\medskip}
I & II & 15 & 0.5 & 1 & 4 & 1 & 0.5 & 0.08 & 0.1\\
II & I & 0.36 & 0.5 & 1 & 4 & 15 & 1 & 0.1 & 0.1\\
III & II & 1 & 0.5 & 5 & 2 & 1 & 2 & 1 & 0.1\\
IV & I & 2 & 0.5 & 0.1 & 2 & 1 & 1 & 5 & 10\\
\noalign{\smallskip} \hline
\end{tabularx}
\end{table}

\begin{figure*}[t]
\begin{center}
\includegraphics[width=13cm]{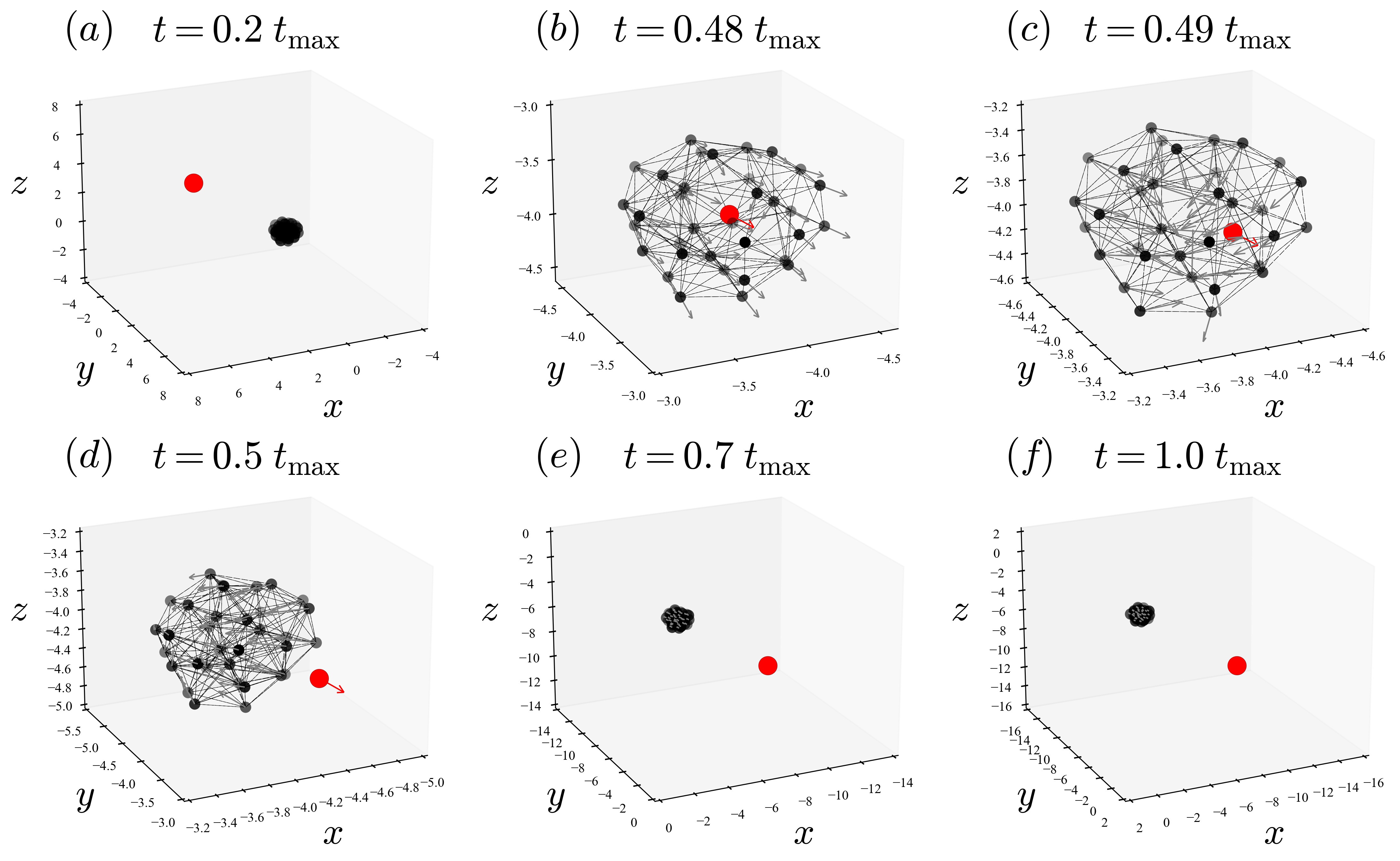}
\end{center}
\caption{
The results of 3D simulation for Pattern I: Split and Reunion. The images demonstrate the behavior of the schooling prey and the predator, as well as the associated condition of $\epsilon$-school at: ($a$) $t = 0.2 t_{\textrm{max}}$, ($b$) $t = 0.48 t_{\textrm{max}}$, ($c$) $t = 0.49 t_{\textrm{max}}$, ($d$) $t = 0.5 t_{\textrm{max}}$, ($e$) $t = 0.7 t_{\textrm{max}}$, and ($f$) $t = t_{\textrm{max}}$, respectively.
} 
\label{fig07}
\end{figure*}

\begin{figure*}[t]
\begin{center}
\includegraphics[width=13cm]{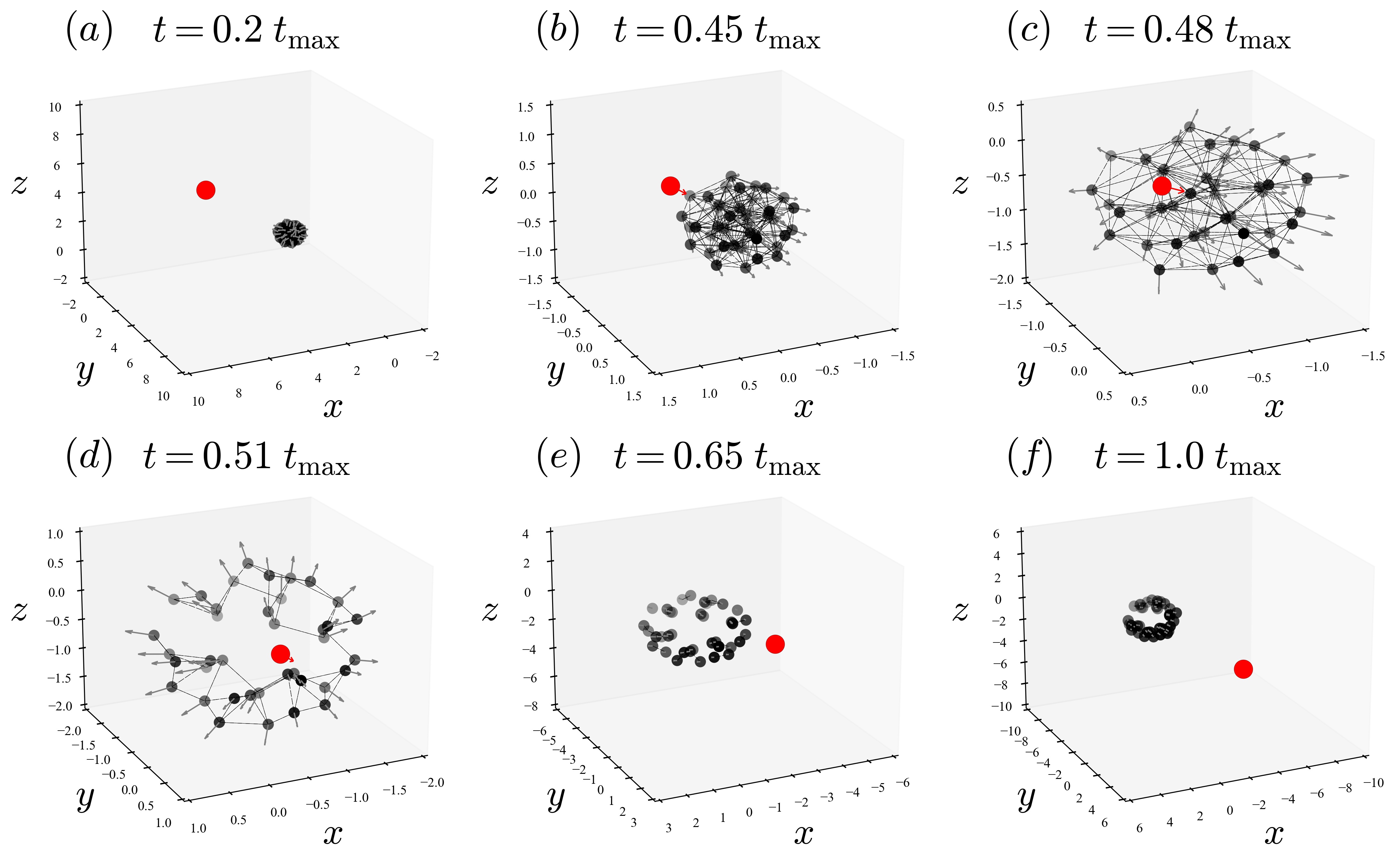}
\end{center}
\caption{
The results of 3D simulation for Pattern II: Split and Separate into Two Groups. The images demonstrate the behavior of the schooling prey and the predator, as well as the associated condition of $\epsilon$-school at: ($a$) $t = 0.2 t_{\textrm{max}}$, ($b$) $t = 0.45 t_{\textrm{max}}$, ($c$) $t = 0.48 t_{\textrm{max}}$, ($d$) $t = 0.51 t_{\textrm{max}}$, ($e$) $t = 0.65 t_{\textrm{max}}$, and ($f$) $t = t_{\textrm{max}}$, respectively.
}
\label{fig08}
\end{figure*}

\begin{figure*}[t]
\begin{center}
\includegraphics[width=13cm]{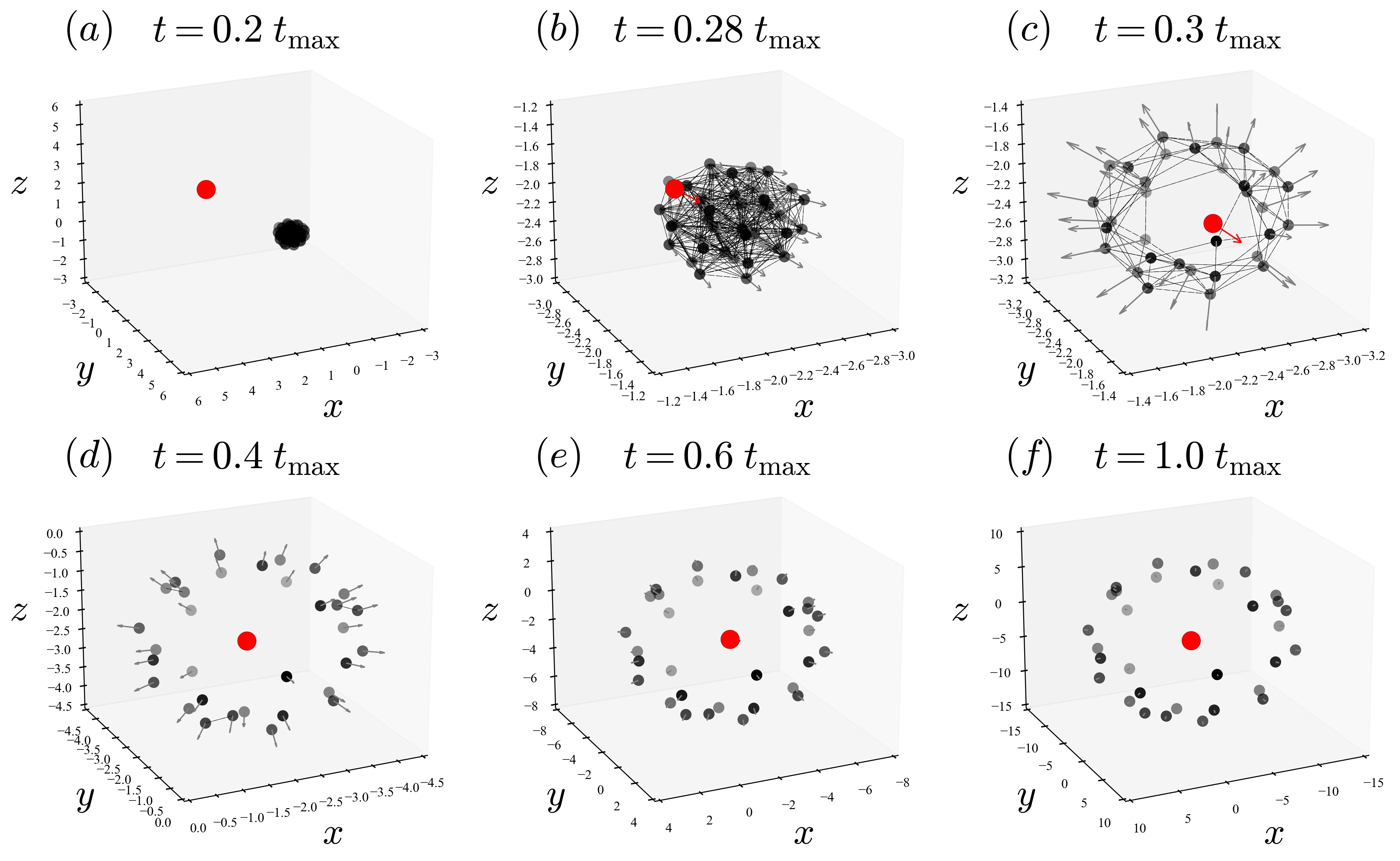}
\end{center}
\caption{
The results of 3D simulation for Pattern III: Scattered. The images demonstrate the behavior of the schooling prey and the predator, as well as the associated condition of $\epsilon$-school at: ($a$) $t = 0.2 t_{\textrm{max}}$, ($b$) $t = 0.28 t_{\textrm{max}}$, ($c$) $t = 0.3 t_{\textrm{max}}$, ($d$) $t = 0.4 t_{\textrm{max}}$, ($e$) $t = 0.6 t_{\textrm{max}}$, and ($f$) $t = t_{\textrm{max}}$, respectively.
}
\label{fig09}
\end{figure*}

\begin{figure*}[t]
\begin{center}
\includegraphics[width=13cm]{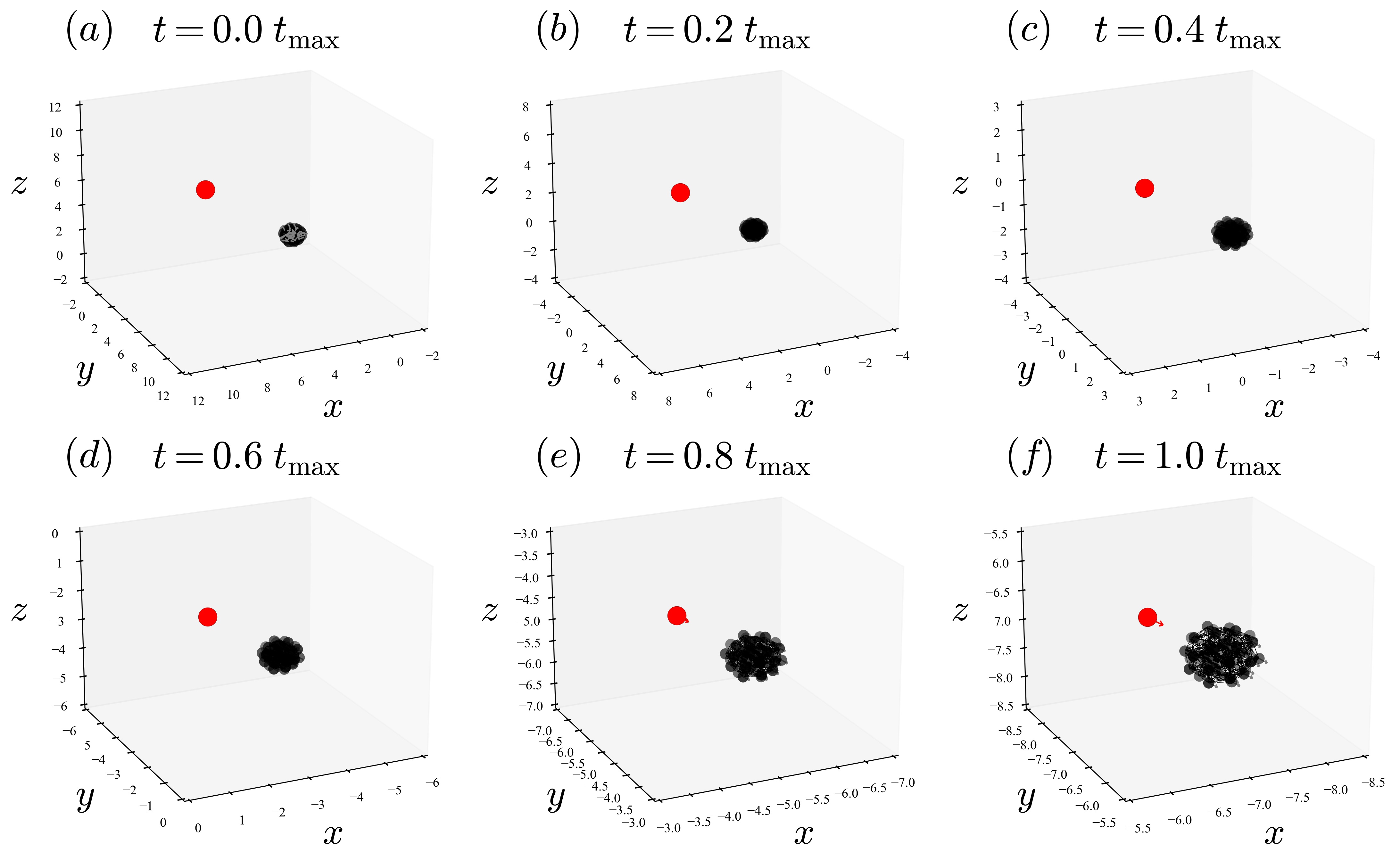}
\end{center}
\caption{
The results of 3D simulation for Pattern IV: Maintain Formation and Distance. The images demonstrate the behavior of the schooling prey and the predator, as well as the associated condition of $\epsilon$-school at: ($a$) $t = 0$, ($b$) $t = 0.2 t_{\textrm{max}}$, ($c$) $t = 0.4 t_{\textrm{max}}$, ($d$) $t = 0.6 t_{\textrm{max}}$, ($e$) $t = 0.8 t_{\textrm{max}}$, and ($f$) $t = t_{\textrm{max}}$, respectively.}
\label{fig10}
\end{figure*}

Figs.~\ref{fig07} - \ref{fig10} exhibit the three-dimensional simulation results for Pattern I (Split and Reunion), Pattern II (Split and Separate into Two Groups), Pattern III (Scattered), and Pattern IV (Maintain Formation and Distance), respectively. In general, the main characteristics of $\epsilon$-schools for all the patterns are similar with the ones observed in the two-dimensional cases. 

\begin{figure}[H]
\begin{center}
\includegraphics[width=8cm]{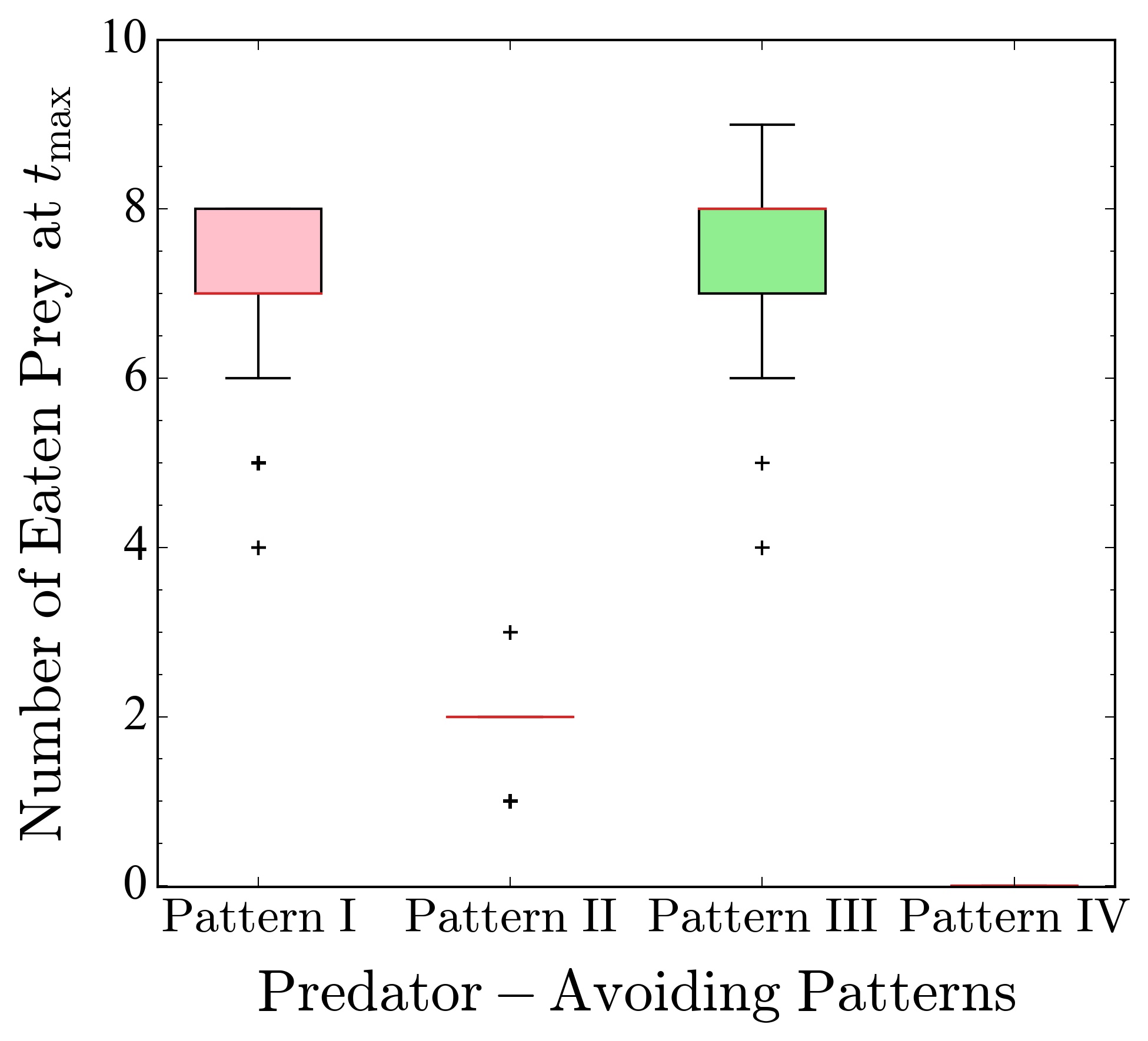}
\end{center}
\caption{Total number of eaten prey for 3D simulation.}
\label{fig11}
\end{figure}

A distinctive feature with the former two-dimensional cases, however, lies in the fact that in the three-dimensional spaces, the individual prey has more spatial flexibility (more degrees of freedom) in its movement to avoid the approaching predator. This reflects in the fewer prey that is being eaten by the predator for each of the predator-avoidance patterns than in their corresponding two-dimensional counterparts. Fig.~\ref{fig11} shows the total number of eaten prey for the four predator-avoidance patterns over 1,000 simulation runs for each of the corresponding patterns. A comparison of Fig.~\ref{fig06} and Fig.~\ref{fig11} supports the erstwhile exposition: fewer prey is being eaten in the three-dimensional cases than the corresponding two-dimensional counterparts due to the higher degrees of freedom in the spatial movements of each individual prey.  

\begin{figure}[H]
\begin{center}
\includegraphics[width=8cm]{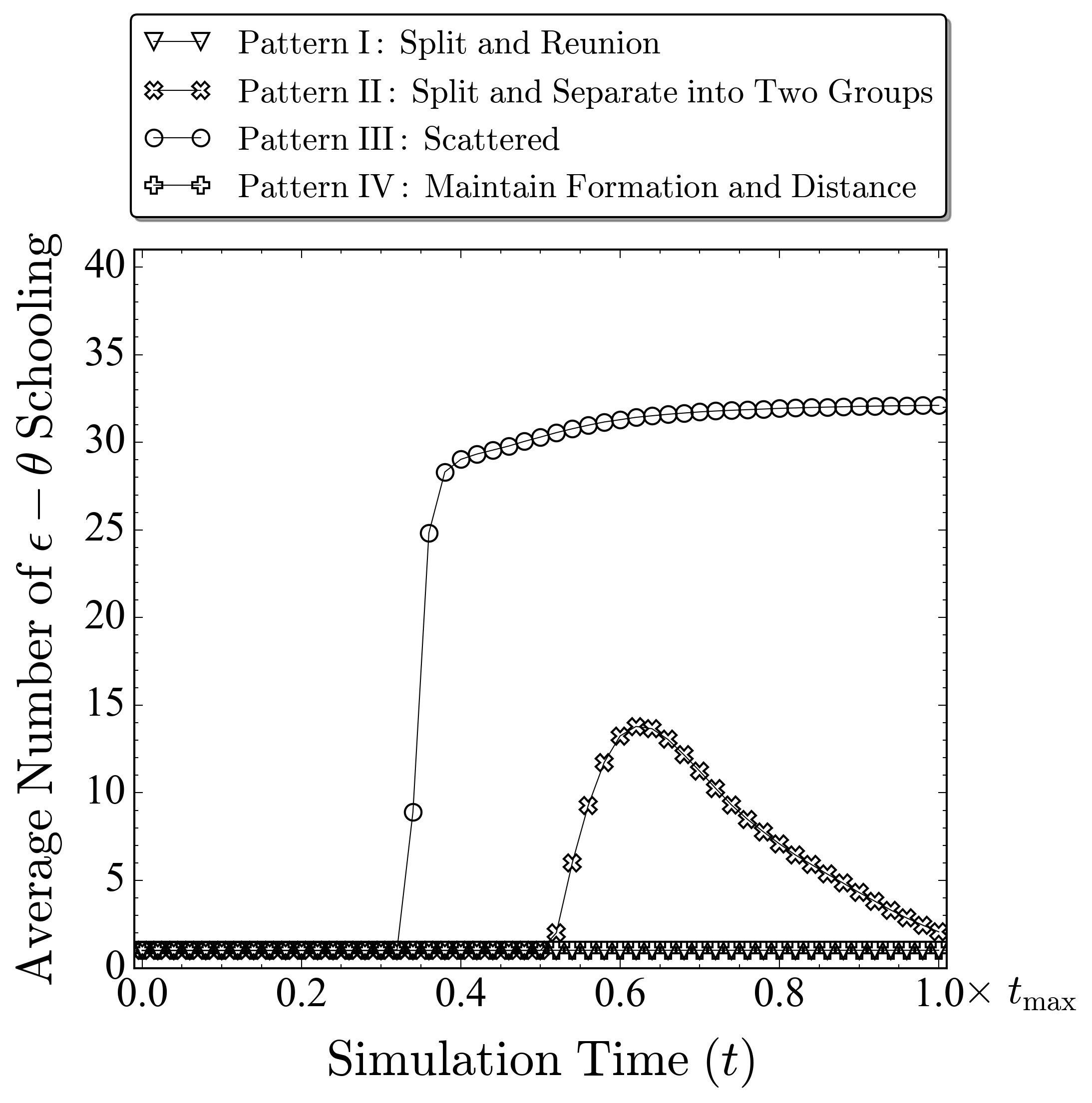}
\end{center}
\caption{Number of $\epsilon$-schools for 3D simulation.}
\label{fig12}
\end{figure}

Fig.~\ref{fig12} demonstrates the number of $\epsilon$-schools for the three-dimensional cases. Here again, we can observe that the $\epsilon$-schools structure for all of the associated predator-avoidance patterns exhibit relatively similar characteristics with their respective two-dimensional cases. Such a consistent result between the two- and three-dimensional simulations reflects the reliability and robustness of our model \eqref{eq01} in describing the transformation of geometrical structure of the schooling prey during predation threat of a solitary predator.

%%% SECTION 4: CONCLUSIONS
\section{Conclusions}
As a final remark, this paper extends the study of the stochastic differential equation model of predator-avoidance in fish schools as presented in Ref.~\citen{Hartono2022Stochastic}. We proposed a concept of $\epsilon$-school as a mathematical representation of the geometrical structure of the schooling prey fish.

By analyzing four different predator-avoidance patterns in both two and three-dimensional spaces, we found that the number of $\epsilon$-schools varies dynamically during the predator's approach. Generally, in the first two patterns, we observed an initial increase in the number of $\epsilon$-schools followed by a decrease to either one or two structures. Pattern III, however, exhibits a distinct characteristic in which the number of $\epsilon$-schools continued to increase until the end of the simulation. A constant unitary $\epsilon$-school is found in Pattern IV. These results suggest that when a predator approaches, individual fish in the school prioritize their attention to the predator rather than maintaining their formation with other schoolmates. Such a finding is consistent with real-life behavior of schooling fish in the natural aquatic ecosystem.

The results of this study contribute to a better understanding of the collective behavior of fish schools and can potentially have implications for the study of animal behavior and group dynamics in various species. Further research can be conducted to explore the impact of various parameters, such as the value of $\epsilon$, as well as the intensity of the noise, on the observed patterns.

%%% ACKNOWLEDGMENTS
\section*{Acknowledgments}
The work of the first (A.D.H.) and the second (T.V.T.) authors were supported by JSPS KAKENHI Grant Number 19K14555.

%%% COMPETING INTERESTS
\section*{Competing Interests}
The authors declare no competing interest in this study.

%%% CONTRIBUTION
\section*{Contribution}
In this study, the first author (A.D.H.) contributes upon developing the computer codes used for the numerical simulations as well as writing the initial draft. The second (T.V.T.) and the last (L.T.H.N.) authors contribute upon designing and conceptualizing the mathematical model as well as revising the initial manuscript.

%%% REFERENCES USING BIBTEX
\bibliographystyle{ws-ijmi}
\bibliography{List_of_References}

\end{multicols}

\end{document}